\documentclass[lettersize,journal]{IEEEtran}
\usepackage{amsmath,amsfonts,amssymb}
\usepackage{algorithm}
\usepackage{algpseudocode}
\usepackage{array}
\usepackage[caption=false,font=normalsize,labelfont=sf,textfont=sf]{subfig}
\usepackage{textcomp}
\usepackage{stfloats}
\usepackage{url}
\usepackage{verbatim}
\usepackage{graphicx}
\usepackage{multirow}
\usepackage{cite}
\usepackage[font=small]{caption}
\usepackage{xcolor}
\usepackage{hyperref}
\usepackage{dsfont}
\usepackage{mathtools}
\usepackage{threeparttable}
\usepackage{booktabs}
\begin{document}
	
\title{Situationally Aware Rolling Horizon Multi-Tier\\ Load Restoration Considering Behind-The-Meter DER}

%\title{Coordinated Multi-Tier Load Restoration Under Highly Congested Distribution Systems With Behind-the-Meter DERs Situational Awareness}

\author{IEEE Publication Technology,~\IEEEmembership{Staff,~IEEE,}
	% <-this % stops a space
	%\thanks{This paper was produced by the IEEE Publication Technology Group. They are in Piscataway, NJ.}% <-this % stops a space
	%\thanks{Manuscript received April 19, 2021; revised August 16, 2021.}
}

\author{Wenlong~Shi,~\IEEEmembership{Member,~IEEE},~Junyuan~Zheng,~\IEEEmembership{Student Member,~IEEE},~and~Zhaoyu~Wang,~\IEEEmembership{Senior  Member,~IEEE}     \thanks{This work was partially supported by the Power System Engineering and Research Center under Grant PSERC S-110, the U.S. Department of Energy Office of Energy Efficiency and Renewable Energy under Grant DE-EE0011234, and the National Science Foundation under Grant ECCS 2042314.}  
	\thanks{The authors are with the Department of Electrical and Computer Engineering,
		Iowa State University, Ames, IA 50011 USA (e-mail:
		wshi5@iastate.edu; zhengjy@iastate.edu; wzy@iastate.edu).}% <-this % stops a space
\thanks{(Corresponding author: Zhaoyu Wang.)}
}



\setlength{\abovedisplayskip}{3.5pt}
\setlength{\belowdisplayskip}{3.5pt}

\maketitle

\begin{abstract}

Restoration in power distribution systems (PDSs) is well studied, however, most existing research focuses on network partition and microgrid formation, where load transfer is limited to adjacent feeders. This focus is not practical, as when adjacent feeders lack sufficient capacity, utilities may request support from more distant feeders in practice. Such a hirarchical restoration is complex, especially when involving changing system conditions due to cold load pickup and delayed reconnection of behind-the-meter DERs. To fill this research gap, a situationally aware multi-tier load restoration framework is proposed. Specifically, models are proposed to describe the multi-tier load restoration, including the multi-tier load transfer and substation transformer and feeder protection  models. By introducing binary actional switching variables and load block transfer variables, the models effectively captures the dynamics of switches and multi-tier transfer process. To integrate situational awareness of evolving system conditions, the problem is formulated as a mixed-integer linear program (MILP) and then embedded within a rolling horizon optimization. Particularly, a set of safeguarded constraints are developed based on segment-level restoration reward bounds to mitigate the myopia of traditional rolling horizon optimization. The proposed safeguarded rolling strategy guarantees that each time step is lower bounded by a $(1-\varepsilon)$-fraction of its optimal restoration potential, thereby balancing short-term switching decisions with long-term restoration goals. Finally, cases studies on the modified IEEE 123-node test feeder validate the proposed multi-tier restoration framework.

\end{abstract}

\begin{IEEEkeywords}
	
	Behind-the-meter, cold load pickup, DER delayed reconnection, multi-tier load transfer, multi-tier restoration, overloading, rolling optimization,  situational awareness.

\end{IEEEkeywords}

\section*{Nomenclature}

\noindent\textit{A. Parameters}

\begin{description}[\IEEEsetlabelwidth{$(\cdot)_r + \textbf{i}(\cdot)_i$}]
	\item[$P_T^\text{max}$] Rated active power capacity of transformer $T$.
	\item[$Q_T^\text{max}$] Rated reactive power capacity of transformer $T$.
	\item[$r_{ij}$] Resistance of distribution line $(i,j)$.
	\item[$x_{ij}$] Reactance of distribution line $(i,j)$.	
	\item[${V}^\text{min}$] Minimum nodal voltage limit.
	\item[${V}^\text{max}$] Maximum nodal voltage limit.
	\item[$S_j$] Load surge factor considering CLPU.
	%\item[$I_e^\text{max}$] Picking up current of fuse $e$.
	\item[$p_i$] Load demand of node $i$.

\end{description}

\noindent\textit{B. Variables}

\begin{description}[\IEEEsetlabelwidth{$(\cdot)_r + \textbf{i}(\cdot)_i$}]
	\item[$D_b^t$] Aggregated demand of load block $b$ at step $t$.
	\item[$v_{bf}^t$] Load block status indicator, 1 if load block $b$ is supplied by feeder  $f$ at step $t$. 
	\item[$u_{if}^t$] Binary, if the sum over all $f$ equals 0, it means the load at node $i$ is shed or not restored.
	\item[$P_f^t$] Active power flow into feeder $f$ at step $t$.
	\item[$Q_f^t$] Reactive power flow into feeder $f$ at step $t$.
	\item[$\Delta o_s^t$] Binary, 1 if implementing open operation on switch $s$ at step $t$, otherwise 0.
	\item[$\Delta c_s^t$] Binary, 1 if implementing close operation on switch $s$ at step $t$, otherwise 0.
	\item[$\Delta v_{bf}^t$] Binary dynamic load transfer variable, 1 if load block $b$ is transferred to feeder $f$ at time  $t$.
	\item[$\gamma_s^t$] Binary switch status indicator, 1 if switch $s$ is connected at step $t$, otherwise 0.

\end{description}

\noindent\textit{C. Indices and Sets}

\begin{description}[\IEEEsetlabelwidth{$(\cdot)_r + \textbf{i}(\cdot)_i$}]
    
    \item[$f,f'$] Index of feeders,  $f'$ is an adjacent feeder of $f$.
    \item[$t\in\mathcal{T}$] Index of restoration time step.
    \item[$\mathds{F}_b$] Set of  adjacent feeders that load block $b$ can be possibly transferred.
    \item[$\mathds{F}_T$] Set of feeders supplied by transformer $T$.
    \item[$\mathds{L}_f$] Set of line segments of feeder $f$.
    \item[$\xi\in \Xi$] Index of situationally aware system state.
    \item[$T\in\mathds{T}$] Index of substation transformers.
    \item[$\mathds{C}_{if}$] Set of child node of $i$ with respect to feeder $f$.
    
    \item[$\mathds{C}_{b'f}$] Set of child block of $b'$ relative to feeder $f$.
    \item[$\mathds{S}_{b'b}$] Set of switches between load blocks $b'$ and $b$.

\end{description}

\section{Introduction}

\IEEEPARstart{R}{estoration} in power distribution systems (PDSs) is a longstanding and critical concern both in academia and industry. Faults caused by overloads, transformer failures, vegetation contact, or weather-induced damage frequently disrupt power delivery, leading to service interruptions. When a fault occurs, the protection system is the first to respond \cite{sati2021optimal}. A well-coordinated protection scheme attempts reclosing to determine whether the fault is transient. If the fault is permanent, switch operation will be executed to isolate the fault. This isolation is typically performed from both upstream and downstream of the fault. Hence, the unaffected islanded areas can be back-fed by other feeders through tie lines for restoration \cite{jooshaki2021milp,wang2015decentralized}.

In literature, research regarding PDS restoration can broadly be divided into two classes: restoration in passive feeders, and restoration with distributed energy resource (DER) integration. For passive feeders, the research mainly focuses on identifying the optimal switching operations involving sectionalizing and tie switches to transfer unserved load to backup feeders. Typically, the problem is formulated as a network reconfiguration task, with objectives that include maximizing the total restored load \cite{chen2009quantitative}, minimizing the number and cost of switching actions \cite{ghofrani2014distribution}, reducing the restoration time \cite{cebrian2016restoration,jooshaki2020reliability}, and improving load balance across feeders \cite{ke2004distribution}. It should satisfy a set of operational constraints, which are imposed on radiality, power balance, voltage drop, and thermal capacity. In contrast, restoration with DERs extend beyond only network topology reconfiguration to include the coordinated utilization of DERs. As a result, recent studies have shifted toward emerging topics, such as dynamic microgrid formation, and co-dispatch of DERs and substations. These studies introduce new tasks, such as dynamic microgrid formation and the co-dispatch of DERs and substation power, with additional objectives considered, such as minimizing load shedding \cite{zhao2024transportable,chen2017multi,abbasi2017parallel}, reducing DER operational costs \cite{gilani2022microgrid,li2023restoration,vargas2021optimal}, and improving resilience indices and energy efficiency \cite{wang2018risk,ye2020resilient,wang2019coordinating,liu2022bi}. Except constraints for passive feeders, restoration with DERs should model the dynamics of microgrid expansion and the process of energy generation and transition. In particular, as the integration of inverter based resources, the system low inertial feature resulted by inverter limits should be addressed. For example, control strategies and transient dynamics can be included as closed form constraints in the problem to enhance system stability during restoration \cite{liu2023utilizing,10700608,10858306}.

PDSs, particularly in urban areas, have become increasingly congested compared to the past. This trend is primarily driven by widespread electrification, with electric vehicle (EV) charging and the adoption of electric heat pumps \cite{shao2022generalized}.  Subsequently, a fundamental question raises — ``are there still untapped opportunities to further enhance restoration performance beyond the limitations of fixed mindset?" We argue that the answer is yes. Based on the literature review, we observe that switching operations form the foundation of PDS restoration \cite{wang2015self}. Their operations enable fault isolation, define microgrid boundaries, and facilitate the transfer of unserved load to healthy sections. Nonetheless, a common limitation is the implicit assumption that each islanded load block can only be restored by utilizing the capacity of an adjacent feeder. If the substation transformer supplying that adjacent feeder lacks sufficient spare capacities, the restoration process halts, rendering the solution infeasible. Such an assumption is not practical, since utilities often adopt a more flexible and hierarchical approach, examining not only the immediate adjacent feeder but also more distant feeders for available capacity \cite{zidan2012cooperative}. If such capacity can be found, a cascaded load transfer is implemented, until sufficient capacity is freed on the immediate adjacent feeder to accommodate the islanded load block. However, this restoraion process is typically guided by expert-based rules, and remains unaddressed from a formal scientific perspective.

We want to emphasize that the problem being studied is not simply a network partitioning as in \cite{chen2015resilient}, or a feeder load balancing as in \cite{gao2021multi}. While considering substation transformers, feeders, DERs and loads collectively in the context of cascaded restoration process, their complicated  inter-dependencies and interactions present significant challenges:

\begin{enumerate}
	
	\item Component Overloading: Restoring islanded load blocks through other healthy feeders requires serious attention. The  additional loads can be  substantial, in particular if cold load pickup (CLPU) phenomenon is involved, and overstress key components such as substation transformers and conductors. For example, a survey performed by IEEE involving 102 North American utilities reported that at least 87\% of them experienced overloading when attempting to restore loads after an extended outage \cite{instance1290}.

	\item Real-Time Load Difference: When restoring an islanded load block, the actual load demand is unobservable since all loads are de-energized, and the AMI meters may be offline. As a result, the restoration solution can only rely on worst-case peak load estimates as known parameters to ensure an effective load transfer, without overloading any transformer capacities. Nonetheless, this assumption often leads to under-utilization of transformer capability, because the actual load demand is mostly lower than the worst peak, especially when considering CLPU.
	
	\item Delayed DER Reconnection:  During outages, DERs, in particular the behind-the-meter (BTM) units, are tripped offline by design in accordance with standard protection protocols. Their reconnection will be delayed until the network becomes stable. Thus, after restoration, a higher net load will be observed than under normal conditions. However, at an uncertain point in time, these DERs may automatically reconnect, resulting in a sudden reduction in net load. These two opposing behaviors introduce significant uncertainty into the restoration process, because it dynamically alters transformer loading conditions.
	
\end{enumerate}

To address the challenges, we propose a situationally aware multi-tier load restoration framework, executing the restoration in a sequential manner. At each step, one islanded load block is transferred for restoration. After each transfer, the net load is measured and aggregated through AMI meters, which reflects both the actual load consumption and the delayed behind-the-meter DER reconnection. By accumulating these information over time, the restoration strategy can be dynamically adjusted to prevent potential overloading and fully exploit the capacity of substation transformers. Main contributions are as follows:

\begin{itemize}
	\item Mathematical models with constraints are developed to support the situationally aware multi-tier load restoration framework. First of all, a multi-tier load transfer model is proposed to capture sequential switching actions. By introducing binary dynamic actional switching and load block transfer variables, the model effectively represents the dynamic behavior of switches and restoration, which cannot be captured through traditional static switch and load block status variables. 
	Second, transformer capacity limits and feeder overloading prevention constraints are incorporated considering CLPU effects. The overloading challenges imposed by an evolving multi-tier topology are addressed by a unified modeling that links transformer, feeder, and load behaviors. Third, situational awareness is embedded as evolving system states by accounting for both the actual net demand and  delayed reconnection of behind-the-meter DERs. The state representation enables informed  real-time decision-making, hence enhancing the efficiency and adaptability of the restoration.
	
	\item The situationally aware multi-tier load restoration problem is formulated as a mixed-integer linear programming (MILP) model embedded in a rolling horizon framework. By sliding a rolling window one step forward at a time, the optimization  dynamically incorporates real-time data collected from AMI, capturing actual net load that reflects both true consumption and behind-the-meter DER reconnections. In addition, to mitigate the effects of myopic solutions resulted by time-coupled switching actions, we introduce a set of safeguarded constraints. Specifically, the rolling horizon is partitioned into several blocks, and for each block, we impose safeguarded constraints ensuring that the cumulative restoration reward never falls below a 
	$(1-\varepsilon)$-fraction of the block's optimal restoration potential. The proposed algorithm effectively balances short-term decisions with long-term restoration goals.
	
\end{itemize}

The remainder of this paper is organized as follows. Section III presents the models, including load transfer model, transformer  and feeder protection model, and situational awareness from AMI meters.
In Section IV, the problem is formulated as a rolling horizon optimization with safeguarded constraints. Section V presents simulation results to demonstrate the the proposed framework. Finally, Section VI concludes the paper and discusses future research directions.

\vspace{-5pt}

\section{Situationally Aware Multi-Tier Load Restoration   Description}

\begin{figure}[t]
	\centering
	\includegraphics[width=3.4in]{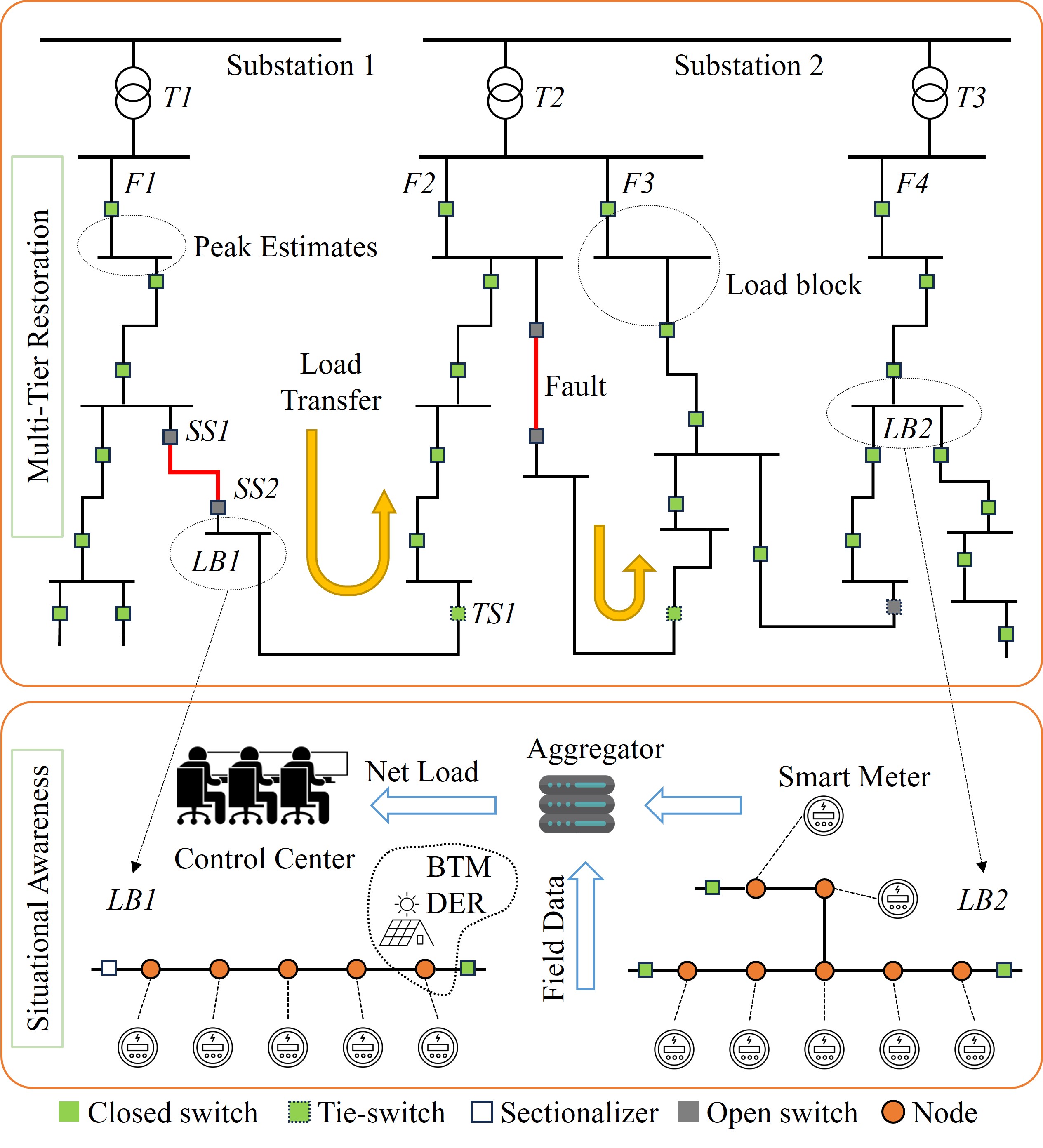}
	% where an .eps filename suffix will be assumed under latex, 
	% and a .pdf suffix will be assumed for pdflatex; or what has been declared
	% via \DeclareGraphicsExtensions.
	%\vspace{-5pt}
	\caption{An illustration of rolling horizon multi-tier load restoration with situational awareness of evolving system state.}
	\label{Model}%\vspace{-5pt}
\end{figure}

In this paper, a multi-tier load restoration is studied leveraging situational awareness. As shown in Fig. \ref{Model}, the substation consists of an HV bus feeding several step‐down transformers. Each transformer $T\in\mathds{T}$ then serves one or more distribution feeders, indexed by $f\in\mathds{F}_T$. For example, as shown in Fig. \ref{Model}, the two substations contains three transformers, each supplying one, two, and one feeders, respectively. Also, we use $i,j\in\mathds{N}$ for denoting nodes, and $(i,j)\in\mathds{L}$ for lines. If there is a fault occurring on line $(i,j)$, the system is equipped with switches to isolate the fault from both upstream and downstream, and transfer the downstream load to other feeders. Herein, we use $s\in\mathds{S}$ to represent the set of switches including sectionalizing switches and tie-switches. In Fig. \ref{Model}, we can see that {SS1} and {SS2} are opened, and {TS1} is closed such that load block {LB1} can be transferred to feeder {F2}. This restoration process is the so-called load transfer, which is very typical in practice.

However, a challenge will arise when the network is highly congested. For example, if transformer {T2} where feeder {F2} is connected lacks capacity to accommodate load block {LB1}, a multi-tier restoration mechanism will be triggered to achieve a coordinated load restoration across substation transformers and distribution feeders. Here, a tier is defined as the total number of successive tie‐switch hops needed to complete a successful load transfer. For example, if {LB1} can be restored by closing two tie-switches sequentially, the restoration is classified as a tier-2 restoration. A hop is a single decision time step, which contains a coordinated pair of switch operations, one opening and one closing, to transfer a load block from its current feeder to an adjacent feeder. For example, in Fig. \ref{Model}, a step $t$ contains opening {SS3} and closing {TS1} for load block transferring {LB1}. In addition, a load block $b\in\mathds{B}$ is the smallest load area that can be isolated by switches. In other words, all loads within a load block are hardwired together, thereby will be transferred to other feeders together without individual load-level control.

Another issue exists in the uncertainties of load demand and the delayed reconnection of behind-the-meter DERs. Restoration planning is derived based on peak load estimates, which can deviate from actual consumption after restoration. Behind-the-meter DERs add further complexity, as they are not utility-owned and cannot be directly controlled. During outages, these DERs are automatically tripped offline, but will reconnect at some time, which cause a sudden drop in net load. Given the uncertainties, situational awareness becomes critically important.  Real-time monitoring of net demand through AMI meters enables system operators to make informed decisions based on actual operating conditions. For example, as shown in Fig. \ref{Model}, field data from smart meters and aggregators are continuously collected and transmitted to the control center. This allows the operator to capture the system state and adaptively determine switching actions in a sequential and responsive manner.

\vspace{-5pt}

\section{Modeling Multi-Tier Load Restoration With Situational Awareness}\label{MDS}

In this section, we propose the models which constitute the situationally aware multi-tier load restoration framework. The models capture the dynamics of multi-tier load transfer, as well as the integration of  situational awareness, particularly considering the delayed reconnection of behind-the-meter DERs. 
In addition, operational constraints are developed such as power flow, transformer and feeder protection under CLPU, as well as multi-tier topology reconfiguration. In the rest  of this section, these models are presented and discussed in detail.

\vspace{-5pt}

\subsection{Multi-Tier Load Transfer Model}

\subsubsection{Load Block Model} A load block refers to the minimum unit that can be transferred between feeders using one step of restoration. It is important to clarify the multi-tier load transfer does not refer to physically connecting an islanded load block directly to a higher-tier or distant feeder. Instead, it employs a sequential switching operations such that, at each step, transfer downstream load blocks from one feeder to an adjacent feeder, freeing up transformer tier by tier. Once sufficient capacity is released, the islanded load block can then be picked up by its nearest tier-1 feeder. Also, each load block $b$ has a set of tier-1 feeders, represented by $\mathds{F}_b$, indicating that it can be transferred only to one of these adjacent feeders. Accordingly, we use $z_{bf}^t$ to denote the binary load block status indicator of block $b$ with respect to feeder $f$, where $f\in\mathds{F}_b$, at step $t$. Then, to ensure that each load block can be supplied by at most one feeder, the following constraint is applied:
\begin{eqnarray}\label{MLTM1}
	\textstyle\sum\nolimits_{f\in\mathds{F}_b}z_{bf}^t\leq 1,\forall b,t.
\end{eqnarray}

Also, a downstream load block $b'$ of $b$ with respect to feeder $f$ can be picked up by feeder $f$ only when $b$ is already picked up by $f$. The corresponding constraint is stated as
\begin{eqnarray}\label{MLTM1}
	z_{b'f}^t\leq z_{bf}^t,\forall b,f,t.
\end{eqnarray}

Furthermore, if a load block is supplied by feeder $f$, all the nodes within it can subsequently be energized:
\begin{eqnarray}\label{MLTM2}
	0\leq u_{if}^t\leq z_{bf}^t,\forall i\in \mathds{N}_b,t.
\end{eqnarray}

Constraint (\ref{MLTM2}) also implies that node $i$ can never be supplied by feeder $f$, when its load block does not belong to feeder $f$. In addition, to capture the incremental characteristic of multi-tier load restoration, we introduce dynamic actional load block transfer variable $\Delta z_{bf}^t\in\{0,1\}$. When $\Delta z_{bf}^t=1$, it indicates that load block $b$ is restored at step $t$ by feeder $f$. This dynamic incremental restored load can be given by
\begin{eqnarray}\label{MLTM21}
	\mathcal{D}_b^t=D_b^t\sum\nolimits_{f\in\mathds F_b}\Delta z_{bf}^t, \forall b,t.
\end{eqnarray}

Equation (\ref{MLTM21}) introduces nonlinearity when included as constraints into the optimization problem. The related linearization technique is presented in Section \ref{PFLT}. Also, to ensure that the load block that is restored in  previous steps is not redundantly counted, we impose the following constraint:
\begin{eqnarray}\label{MLTM23}
	\Delta z_{bf}^t \leq 1-\sum\nolimits_{f\in\mathds{F}_b}z_{bf}^{t-1}, \forall b,t,
\end{eqnarray}
\begin{eqnarray}\label{MLTM24}
	 z_{bf}^t \geq \Delta z_{bf}^{t}, \forall b,t,f.
\end{eqnarray}

Constraint (\ref{MLTM23}) represents that transferring load block that is already energized is not considered as restoration. Constraint (\ref{MLTM24}) updates the  load block status $z_{bf}^t$ based on the current step dynamic load transfer decision.

\subsubsection{Dynamic Actional Switching Model}

At each step, the load transfer is performed by at most two switching operations, one opening and one closing. Therefore, each restoration step corresponds to one time step. This restriction reflects standard industry practice considering reliability, safety, and protection. In practice, utilities rarely attempt more than two operations in one single restoration step \cite{liu2020comprehensive}. For capturing this operational practice, we introduce for each switch $s$ at time step $t$ two dynamic actional switching variables $\Delta o_s^t,\Delta c_s^t\in\{0,1\}$, where $\Delta o_s^t=1$ denotes an opening action, and $\Delta c_s^t=1$ denotes a closing action. Note that from the perspective of multi-tier restoration, there is no strict distinction between sectionalizing switches and tie-switches. For example, a sectionalizing switch that is previously opened may function as a tie-switch in some future steps, depending on the evolving multi-tier topology. To impose the limitations on the number of switch operations, the following constraints are applied:
\begin{eqnarray}\label{MLTM3}
	\Delta o_s^t+\Delta c_s^t\leq 1, \forall s,t,
\end{eqnarray}
\begin{eqnarray}\label{MLTM4}
	\textstyle\sum\nolimits_{s\in\mathds{S}}\{\Delta o_s^t+\Delta c_s^t\}\leq 2, \forall t.
\end{eqnarray}

Constraint (\ref{MLTM3}) ensures no switches can be both opened and closed at the same time in one step. Constraint (\ref{MLTM4}) ensures an industry‐standard limit of at most two switch actions per step. Based on the actions, the switch status can be updated as
\begin{eqnarray}\label{MLTM5}
	\gamma_s^{t+1}=\gamma_s^t-\Delta o_s^t+\Delta c_s^t, \forall s,t,
\end{eqnarray}
\begin{eqnarray}\label{MLTM6}
	\Delta o_s^t\leq \gamma_s^t, \forall s,t,
\end{eqnarray}
\begin{eqnarray}\label{MLTM7}
	\Delta c_s^t\leq 1- \gamma_s^t, \forall s,t.
\end{eqnarray}

Constraint (\ref{MLTM5}) models the transition of switch status by dynamically capturing the effect of opening and closing actions. Constraints (\ref{MLTM6}) and (\ref{MLTM7}) ensure that one switch can be opened only if it is initially closed, and be closed only if it is initially open. Note that the collective behavior enforced by constraints (\ref{MLTM3})–(\ref{MLTM7}) cannot be represented by using a single static switch status variable, i.e., when only  $\gamma_s^h$ is available.

\subsubsection{Topology Reconfiguration Model} 

To support the multi-tier load transfer, the network topology should be dynamically reconfigured. Then, the following constraints are applied:
\begin{eqnarray}\label{FIC1}
	z_{bf}^t + z_{b'f}^t  \leq 1+\gamma_s^t, \forall s=\mathds{S}_{bb'},f,t,
\end{eqnarray}
\begin{eqnarray}\label{FIC2}
	z_{bf}^t - z_{b'f}^t  \leq 1-\gamma_s^t, \forall s=\mathds{S}_{bb'},f,t,
\end{eqnarray}
\begin{eqnarray}\label{FIC3}
	z_{bf}^t-z_{b'f}^t  \geq -(1-\gamma_s^t), \forall s=\mathds{S}_{bb'},f,t.
\end{eqnarray}

Constraints (\ref{FIC1})-(\ref{FIC3}) ensure that when load blocks $b$ and $b'$ belong to different feeder, the only way to complete the load transfer is closing this switch. In other words, if both blocks $b$ and $b'$ are supplying by feeder $f$, the switch between these two blocks must be closed. In addition, when a fault occurs, Fault Location, Isolation, and Service Restoration (FLISR) systems quickly detect the fault location and isolate the affected area to avoid its propagation. Then, the faulted area is de-energized as a whole that its load block cannot be supplied by any feeder:
\begin{eqnarray}\label{FIC4}
	\textstyle\sum\nolimits_{f\in\mathds{F}_b}v_{bf}^t\leq 0, \forall b\in \mathds{B}_d,t.
\end{eqnarray}

Moreover, all switching actions must be implemented under radiality constraints to avoid the formation of loops during the load transfer process, given by \cite{wang2020radiality}
\begin{eqnarray}\label{FIC5}
	\textstyle\sum\nolimits_{s\in\mathds{S} }\Gamma_s^t = |\mathds{N}| - |\mathds{F}|, \forall t,
\end{eqnarray}
\begin{eqnarray}\label{FIC6}
	\textstyle\sum\nolimits_{b \in \pi_{b'}} F_{b'b}^t - \sum\nolimits_{b \in \pi_{b'}} F_{bb'}^t = -1,  \forall b' \notin \mathds{F},t,
\end{eqnarray}
\begin{eqnarray}\label{FIC7}
	\textstyle\sum\nolimits_{b \in \pi_f} F_{fb}^t  \geq 1,  \forall f \in \mathds{F},t,
\end{eqnarray}
\begin{eqnarray}\label{FIC8}
	-\Gamma_{s}^t|\mathds{N}| \leq F_{bb'}^t\leq \Gamma_{s}^t|\mathds{N}|, \forall s\in\mathds{S}_{bb'},t,
\end{eqnarray}
\begin{eqnarray}\label{FIC9}
	\gamma_{s}^t \leq \Gamma_{s}^t, \forall s\in\mathds{S}_{bb'},t.
\end{eqnarray}

Constraints (\ref{FIC5})-(\ref{FIC9}) enforce the radiality by constructing  a fictitious graph that mirrors the distribution network topology, and applying a single-commodity flow model. Constraint (\ref{FIC5}) decomposes the graph into a forest with $|\mathds{F}|$ disjoint trees, that the sum of connected edges must be $|\mathds{N}| - |\mathds{F}|$. Constraint (\ref{FIC6}) ensures that a non-source node receives exactly one unit of the fictitious commodity, such that the net inflow minus outflow equals $-1$. Constraint (\ref{FIC7}) means that each source node must inject at least one unit of the commodity to initiate connectivity in each tree. Constraint (\ref{FIC8}) enforces the ficticious flow to 0 if the edge is disconnected. Constraint (\ref{FIC9}) links the ficticious graph to distribution network by disconnecting switches.

\vspace{-5pt}

\subsection{Transformer and Feeder Protection Model}

The proposed multi-tier load transfer mechanism can dynamically restore islanded load blocks, while it also introduces changing network topology. To this end, a model is presented to coordinate the effective operation of substation  transformers and distribution feeders. Firstly, the following constraints are applied to prevent transformers from overloading:
\begin{eqnarray}\label{STFM1}
	\sum\nolimits_{f\in\mathds{F}_T} P_f^h\leq P_T^\text{max}, \forall T,h,
\end{eqnarray}
\begin{eqnarray}\label{STFM2}
	\sum\nolimits_{f\in\mathds{F}_T} Q_f^h\leq Q_T^\text{max}, \forall T,h,
\end{eqnarray}
which indicate the total inflow power of all connected feeders must not exceed the transformer $T$ rated capacity. In addition, to prevent feeders from overloading, we have
\begin{eqnarray}\label{STFM112}
	P_f^t\leq P_f^\text{max},Q_f^t\leq Q_f^\text{max}, \forall f,t,
\end{eqnarray}
where $P_f^\text{max}$ and $Q_f^\text{max}$ represent the active and reactive power flow limits of feeder $f$,  determined based on the conductor’s capacity and the pickup settings of downstream protective devices. Constraints (\ref{STFM1})-(\ref{STFM112}) ensure that as load is transferred tier by tier, no components are subjected to the risk of damage. 

Secondly, each feeder $f$ consists of a series of line segments $(i,j)\in\mathds{L}_f$ with line impedance $r_{ij}+jx_{ij}$. To link the feeder level power injection with power flow variables, we have
\begin{eqnarray}\label{STFM3}
	P_f^t=P_{\cdot jf}^{t}, \forall (\cdot,j)\in\mathds{L}_f,t,
\end{eqnarray}
\begin{eqnarray}\label{STFM4}
	Q_f^t=Q_{\cdot jf}^{t}, \forall (\cdot,j)\in\mathds{L}_f,t,
\end{eqnarray}
where $(\cdot,j)\in\mathds{L}_f$ represents the first line segment of feeder $f$. Constraints (\ref{STFM3})-(\ref{STFM4}) indicate that the total real and reactive power delivered by transformer into feeder $f$ should equal the sum of the real and reactive power flows across the first line segment of that feeder. In addition, for each line segment, the linearized DistFlow model is applied \cite{wang2014coordinated}:
\begin{eqnarray}\label{STFM5}
	\textstyle\sum_{j'\in\mathds{C}_{jf}}{P}_{jj'f}^{t}={P}_{ijf}^{t}-{p}_{j}^{t}+\Delta{p}_{ijf}^{t}, \forall (i,j),f,t,
\end{eqnarray}
\begin{eqnarray}\label{STFM6}
	\textstyle\sum_{j'\in\mathds{C}_{jf}}{Q}_{jj'f}^{t}={Q}_{ijf}^{h}-{q}_{j}^{t}+\Delta{q}_{ijf}^{t}, \forall (i,j),f,t,
\end{eqnarray}
\begin{eqnarray}\label{STFM7}
	V_{if}^t-V_{jf}^t= \frac{(r_{ij}{P}_{ijf}^{t}+x_{ij}{Q}_{ijf}^{t})}{V_{0}}+\Delta{v}_{jf}^{t}, \forall j\in\mathds{C}_{if},t.
\end{eqnarray}

Constraints (\ref{STFM5})-(\ref{STFM6}) are used for active and reactive power balance, while constraint (\ref{STFM7}) is for voltage drop calculation. Notice that $\Delta{p}_{ijf}^{t}$, $\Delta{q}_{ijf}^{t}$,  $\Delta{v}_{jf}^{t}$ are slack variables which will be enabled by the following constrains:
\begin{eqnarray}\label{STFM8}
	{\Delta{{ p}}_{ijf}^t}\leq ({1}-z_{b'f}^t){M}, \forall  b'\in\mathds{C}_{bf},(i,j)=\mathds{S}_{bb'},
\end{eqnarray}
\begin{eqnarray}\label{STFM9}
	{\Delta{{q}}_{ijf}^t}\leq ({1}-z_{b'f}^t){M}, \forall  b\in\mathds{C}_{bf},(i,j)=\mathds{S}_{bb'},
\end{eqnarray}
\begin{eqnarray}\label{STFM10}
	{\Delta{{v}}_{jf}^t}\leq ({1}-z_{b'f}^t){M}, \forall j\in\mathds{C}_{if},b'\in\mathds{C}_{bf},(i,j)=\mathds{S}_{bb'},
\end{eqnarray}
\begin{eqnarray}\label{STFM11}
	\Delta{{ p}}_{ijf}^t,\Delta{{ q}}_{ijf}^t,{\Delta{{v}}_{jf}^t}\geq 0, \forall b'\in\mathds{C}_{bf},(i,j)=\mathds{S}_{bb'}.
\end{eqnarray}

Constraints (\ref{STFM8})-(\ref{STFM11}) are used to relax constraints (\ref{STFM5})-(\ref{STFM7}) if load block $b$ is not supplied by distribution feeder $f$ at step $t$. In such cases, the  power flow into block $b$ from transformer $T$ through switch $(i,j)$ is forced to 0 by the following constraint:
\begin{eqnarray}\label{STFM12}
	0\leq P_{ijf}^{t}\leq z_{b'f}^t P_T^\text{max}, \forall  b'\in\mathds{C}_{bf},(i,j)=\mathds{S}_{bb'},f\in\mathds{F}_T,
\end{eqnarray}
\begin{eqnarray}\label{STFM13}
	0\leq Q_{ijf}^{t}\leq z_{b'f}^t Q_T^\text{max}, \forall  b'\in\mathds{C}_{bf},(i,j)=\mathds{S}_{bb'},f\in\mathds{F}_T.
\end{eqnarray}

Note that by enforcing (\ref{STFM1})-(\ref{STFM13}), the load transfer of block $b$ from feeder $f$ to feeder $f'$ (i.e., $z_{bf}^h=0$ and $z_{bf'}^h=1$)  automatically deactivates the DistFlow constraints on the former feeder and activates them on the latter. Accordingly, the feeder injection constraints (\ref{STFM3})-(\ref{STFM4}) and the transformer protection constraints   (\ref{STFM1})-(\ref{STFM2}) are updated to adapt to the new topology whenever an island is re-fed from a different feeder. It ensures that each transformer’s loading is recalculated at each step and kept within its rating no matter how the topology changed. In addition, the nodal voltage drop limits are applied:
\begin{eqnarray}\label{STFM14}
	0\leq V_{if}^t\leq z_{bf}^t{V}^\text{max}, \forall i\in\mathds{N}_b,
\end{eqnarray}
\begin{eqnarray}\label{STFM15}
	\textstyle\sum\nolimits_{f\in\mathds{F}}z_{bf}^t{{V}^\text{min}}\leq\sum\nolimits_{f\in\mathds{F}}V_{if}^t\leq \sum\nolimits_{f\in\mathds{F}}z_{bf}^t{{V}^\text{max}}, \forall i\in\mathds{N}_b,
\end{eqnarray}
which enforce all nodal voltages to 0, if the nodes $i$ in load block $b$ is not supplied by feeder $f$ at time step $t$.

CLPU is another concern during multi-tier load restoration, particularly following extended outages \cite{song2020robust}. CLPU can cause a temporary surge in islanded load blocks, often several times higher than normal levels, due to the simultaneous energization of thermostatically controlled loads, motor inrush currents, and the loss of load diversity. This load surge can persist for hours, overloading the healthy feeders and accelerating the aging of electrical components \cite{gilvanejad2013estimation}. The surge demand can be modeled as $S_jp_j$, with $S_j$ representing the surge factor. For loads in the unaffected  blocks, we have $S_j=1$. Under this consideration, the active and reactive load terms in constraints (\ref{STFM5})–(\ref{STFM6}) are updated to reflect the CLPU accordingly.

\vspace{-5pt}

\subsection{Situational Awareness of System State}
To realize situational awareness, the system state is dynamically updated using real-time field date. Specifically, at each time step $t$, after implementing the switching actions, the load demand of each node is measured by smart meters and transmit back to the control center through AMIs. These measurements capture both the real-time load consumption and delayed DER reconnection after the islanded load blocks are restored. Then, the updated system state $\xi$ can be stated as
\begin{eqnarray}
	\xi_{t+1}
	=\{\gamma^{t+1},\,z^{t+1},p^{t+1}\},
\end{eqnarray}
where $\gamma^{t+1}$ and $z^{t+1}$ are the updated switch statuses and load block energization indicators at $t+1$, respectively, which are deterministically known from the executed actions. Also, $p^{t+1}$ represents the actual load measurements acquired through situational awareness. By embedding these observed information into the system state, the decision-making  becomes responsive to the real-time operating conditions, enhancing the efficiency and adaptability of the multi-tier load restoration. 
\vspace{-5pt}

\section{Situationally Aware Multi-Tier Load Restoration Problem Formulation and Solution}\label{PFLT}

In this section, the proposed multi-tier  restoration problem is formulated as an MILP problem by incorporating the models in section \ref{MDS} as constraints. To enhance situational awareness, particularly in capturing the reconnection of behind-the-meter DERs, the MILP is embedded within a rolling horizon optimization framework. In addition,  to address the myopic issue of traditional rolling approach, a safeguarded rolling strategy is proposed by imposing constraints of restoration reward bounds within each look-ahead segment. In the rest of this section, the  formulation and solution procedure is presented in detail.

\vspace{-5pt}

\subsection{Rolling Horizon Framework}

The proposed coordinated multi-tier load restoration problem is formulated as an MILP  embedded in a rolling horizon optimization framework. In each time step $t$, the updated MILP is solved over a look-ahead rolling window with $\mu$ successive steps. Only the switching decisions corresponding to the first step are implemented to obtain an ``act and observe" principle. After that, the system state is updated by measuring the actual net load by using real-time data from AMI meters. The horizon is then shifted forward by one step, and the process repeats by incorporating both the true load demand and the reconnection of DERs into the next iteration of optimization. Accordingly, the objective function at one time step, aiming to maximize the incremental restored load over $\mu$ steps window, is
\begin{eqnarray}\label{OBJ}
	\max_{\substack{\Delta o,\Delta c}}
	\sum_{t=t'}^{t'+\mu-1}
	\Bigl[\sum_{b\in\mathds B}D_b^t\sum_{f\in\mathds F_b}\Delta z_{bf}^t
	-\alpha\sum_{s\in\mathds S}(\Delta o_s^t+\Delta c_s^t)\Bigr],
\end{eqnarray}
\begin{flalign*}
	&\enspace\mathrm{subject\ to\ constraints\ (1)-(36),\ (39)-(44),}&
\end{flalign*}
where the first term denotes the incremental demand of newly energized blocks at time $t$, and the second term introduces a penalty that imposes a small cost for each pair of switching operation. In sum, the objective function prioritizes switch sequences so that more loads are restored with fewer operations. Notice that the first term is quadratic, which can be linearized by replacing it with $\mathcal{D}^t$ under the following constraints:
\begin{eqnarray}
\textstyle\mathcal{D}^t=\sum\nolimits_{b\in\mathds B}\mathcal{D}_b^t, 
\end{eqnarray}
\begin{eqnarray}
	\Delta v_{bf}^t\leq v_{bf}^t, \forall b,t,
\end{eqnarray}
\begin{eqnarray}
	\Delta v_{bf}^t\leq 1-v_{bf}^{t-1}, \forall b,t,
\end{eqnarray}
\begin{eqnarray}
	\textstyle\mathcal{D}_b^t\leq \sum\nolimits_{i\in\mathds N_b}p_i\sum\nolimits_{f\in\mathds F_b}[u_{if}^t+(1-\Delta v_{bf}^t)M], \forall b,t,
\end{eqnarray}
\begin{eqnarray}
	\textstyle\mathcal{D}_b^t\geq \sum\nolimits_{i\in\mathds N_b}p_i\sum\nolimits_{f\in\mathds F_b}[u_{if}^t-(1-\Delta v_{bf}^t)M], \forall b,t,
\end{eqnarray}
\begin{eqnarray}
	\textstyle-\sum\nolimits_{f\in\mathds F_b}\Delta v_{bf}^tM\leq\mathcal{D}_b^t\leq \sum\nolimits_{f\in\mathds F_b}\Delta v_{bf}^tM, \forall b,t,
\end{eqnarray}
where $p_i$ is the load demand at node $i$. It is a fixed parameter within each iteration, while is updated at the beginning of each new iteration based on observed field data collected from AMI meters. And, $u_{if}^t$ is a load shedding variable, if the load at node $i$ is shed by operating the  smart meter. In addition, for notation simplicity, equation (\ref{OBJ}) is denoted by $\max_x\sum_{t=t'}^{t+\mu-1}f_{\xi}(x)$ in the rest of the paper, where $x$ represents all the variables and $\xi$ denotes the system state, such as network topology and net load demand at time step $t$. Its value represents the maximum restoration reward that can be gained at time $t$.

The introduction of the rolling horizon optimization framework enables the utilization of field data, effectively enhancing situational awareness throughout the multi-tier load restoration process. In contrast, a one-shot MILP formulation are not able to incorporate system uncertainties. Despite their adaptability, rolling horizon approaches also have limitations. In particular, the switching decisions across successive restoration steps are coupled. It means that the current executed action, along with the newly observed system state,  directly prunes the feasible solution space in subsequent iterations. This coupling may trap the multi-tier load restoration in some local optima, prioritizing short-term gains at the expense of long-term potential, such as higher-tier transformer capacity.

\vspace{-5pt}

\begin{figure}[t]
	\centering
	\includegraphics[width=3.4in]{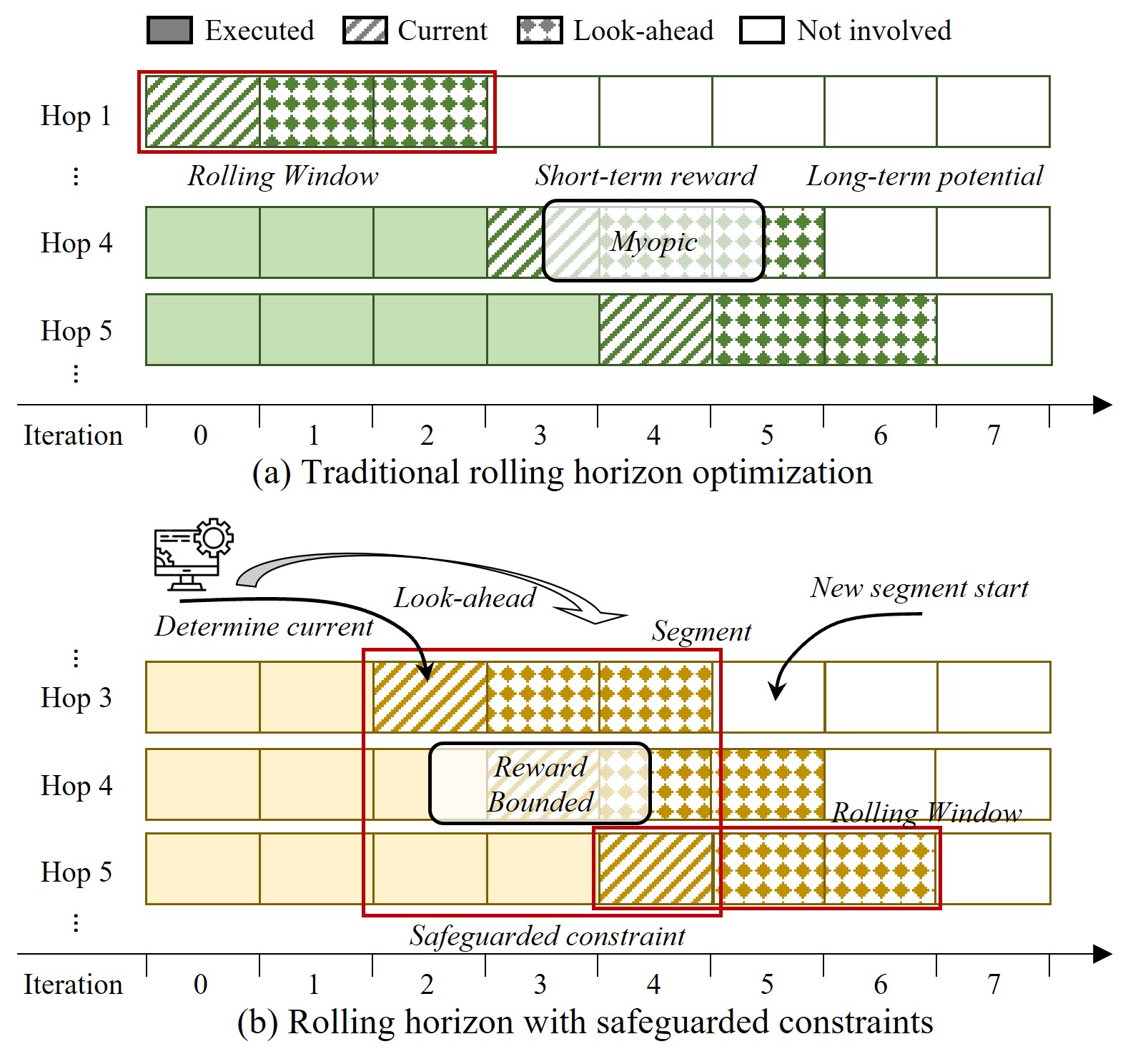}
	% where an .eps filename suffix will be assumed under latex, 
	% and a .pdf suffix will be assumed for pdflatex; or what has been declared
	% via \DeclareGraphicsExtensions.
	%\vspace{-5pt}
	\caption{An illustration of the proposed rolling horizon multi-tier load restoration problem with safeguarded constraints.}
	\label{Safe}\vspace{-10pt}
\end{figure} 

\subsection{Safeguarded Rolling Approach}

In this subsection, safeguarded constraints are proposed to mitigate the myopic associated with traditional rolling horizon approach. Specifically, we begin by presenting a segment‐level performance bound that holds under any start‐state $\xi$, and then integrate it into the problem as safeguarded constraints. These constraints ensure that the optimizer will  lose no more than an \(\varepsilon\)-fraction of 
its optimal restoration potential over the rolling window. Here, a start-state is the system state at the beginning of a given iteration, such as network topology and loading.

\medskip
\noindent\textbf{Theorem 1 (Segment‐Level Reward Bound).} Let $\xi_t$ be the start-state at time step $t$, and $F_t(\xi_t)=\max_x \sum_{t=t'}^{t'+\mu-1}f_{\xi_t}(x)$ be the optimal reward over the next 
$\mu$ steps from start-state  $\xi_t$. Let $F_t^{+}= \max_x\sum_{t=t'}^{t'+\mu-1}f_{\xi}(x)$ be the best possible reward over all feasible start-states $\xi$. For the rolling horizon multi‐tier load restoration problem, there exists a constant $\varepsilon \in[0,1)$ such that, for every iteration $r$, we have
\begin{eqnarray}
	F_t(\xi_t)\geq (1-\varepsilon)F_t^{+}, \forall t,
\end{eqnarray}
where $\varepsilon=A/\Delta D$, with $\Delta D$ denoting the minimum restored load over $\mu$ steps. Also, $A$ is a multiple of $\alpha$, representing the penalty of conducting switching actions over $\mu$ steps.

\medskip

\noindent\textbf{Proof.} Since $\Delta D$ represents the minimum restored load over $\mu$ steps, we have $F_t^{+}\geq \Delta D$. In addition, if we start from the worst start-state, we can always conduct switching operations to restore load blocks at the expense of reward, stated as $F_t^{-}\geq\Delta D-A$. Then, the following inequality holds:
\begin{eqnarray}
	{F_t^{-}}/{F_t^{+}}\geq 1-{A}/{\Delta D},
\end{eqnarray}
by combining with $F_t(\xi_t) \geq F_t^{-}$, the proof is completed. \enspace\enspace  $\blacksquare$

\medskip

Theorem 1 guarantees that, over any successive $\mu$ step look-ahead window, the rolling-horizon optimizer is able to recover at least a $(1-\varepsilon)$-fraction of the maximum restoration reward, even when starting from an adverse state, e.g., highly loaded transformers, offline DERs, or elevated CLPU surges. Herein, $(1-\varepsilon)$ quantifies how close we  remain to the optimal reward, regardless of initial conditions. By selecting a window length $\mu$, a reasonable $\varepsilon$ can always be found. In addition, To extend our segment‐level performance guarantee over the entire whole $|\mathcal{T}|$ steps horizon, we proceed in three steps:

\paragraph{Horizon Partitioning} Based on the $|\mathcal{T}|$ steps horizon, we partition the sequence of steps into $L+1$ non-overlapping segments, which contain one short segment with length $\beta$ and $L$ full segments with length $\mu$. Hence, we have 
\begin{eqnarray}
	|\mathcal{T}|+1 =\beta +L\mu,\ 0\le\beta<\mu,\ L = \left\lfloor {(|\mathcal{T}| + 1)}/{\mu} \right\rfloor.
\end{eqnarray}

The first segment covers steps $\{0,1,\dots,\beta-1\}$, and the next full segments covers steps $ \{\beta+(l-1)\mu,\dots,\beta+l\mu-1\},l\in\{1,2,\dots\}$. Note that the short segment only appears when $|\mathcal{T}|$ can not be divisible by $\mu$. If $\beta=0$, all segments are full.

\paragraph{Segment Maximum} Every time the algorithm enters a new segment, we solve one $\beta$ or $\mu$ steps MILP before rolling, and obtain the best restoration reward of this segment:
\begin{eqnarray}\label{SG11}
	\textstyle F_0^{+}
	= 
	\max 
	\sum\nolimits_{h=0}^{\beta-1} f_{\xi_t}(x_h),
\end{eqnarray}
\begin{eqnarray}\label{SG12}
	\textstyle F_l^+
	=
	\max 
	\sum\nolimits_{h=t}^{t+\mu-1} f_{\xi_t}(x_h),t=\beta+(l-1)\mu.
\end{eqnarray}
where $F_0^{+}$ is for the $\beta$ steps window, and $F_l^{+}$ for the $\mu$ steps window. These parameters provide fixed upper‐bounds on the restoration potential of each segment.

\begin{algorithm}[t]\small
	\caption{\small Safeguarded Rolling Multi‐Tier Restoration}
	\label{ALG}
	\begin{algorithmic}[1]
		\Require Initial state $\xi_0$, horizon $|\mathcal{T}|$, segment length $\mu$.
		\State $\beta = (|\mathcal{T}|+1)\bmod \mu$, $L= \lfloor (|\mathcal{T}|+1)/\mu\rfloor$.
		\For{$l=0$ \textbf{to} $L$}  
		\If{$l=0$}
		\State Solve the MILP, deriving $F_0^{+}
		= 
		\max 
		\sum\nolimits_{h=0}^{\beta-1} f_{\xi_t}(x_h)$.
		\For{$t=0$ \textbf{to} $\beta-1$}		
		\State Solve MILP over steps $[0,\dots,\beta-1]$ with 
		constraints 
		\State $\sum_{h=0}^{\beta-1} f_{\xi_t}(x_h)
		\geq
		(1-\varepsilon)\,F_0^{+}$.
		\State Implement switching $\{\Delta o_s^t,\Delta c_s^t\}$. \State Update system state 
		$\xi_{t+1}\gets\{\gamma^{t+1},v^{t+1},p^{t+1}\}$.
		\EndFor
		\Else
		\State Solve the MILP, deriving the segment reward bound 
		\State $F_l^+
		=
		\max 
		\sum\nolimits_{h=t}^{t+\mu-1} f_{\xi_t}(x_h),t=\beta+(l-1)\mu$.
		\For{$t=[\beta+(l-1)\mu$ \textbf{to} $\beta+l\mu-1$}
		\State Solve MILP over steps $[\beta+(l-1)\mu,\dots,\beta+l\mu-1]$
		\State with constraints $\sum_{h=\beta+(l-1)\mu}^{\beta+l\mu-1} f_{\xi_t}(x_h)
		\geq
		(1-\varepsilon)F_l^+$.
		\State Implement switching $\{\Delta o_s^t,\Delta c_s^t\}$. \State Update system state 
		$\xi_{t+1}\gets\{\gamma^{t+1},v^{t+1},p^{t+1}\}$.
		\EndFor
		\EndIf
		\EndFor
	\end{algorithmic}
\end{algorithm}

\paragraph{Segment Safeguarding} To prevent the optimizer from making myopic decisions, the segment‐level reward bound is leveraged so that, within each segment, the restoration reward can never drop below a $(1-\varepsilon)$-fraction of the segment’s best value. Then, the safeguard constraints are proposed as follows:
\begin{eqnarray}\label{SG1}
	\textstyle\sum_{h=0}^{\beta-1} f_{\xi_t}(x_h)
	\geq
	(1-\varepsilon)\,F_0^{+},
	t=0,1,\dots,\beta-1,
\end{eqnarray}
\begin{eqnarray}\label{SG2}
	\textstyle\sum_{h=\beta+(l-1)\mu}^{\beta+l\mu-1} f_{\xi_t}(x_h)
	\geq
	(1-\varepsilon)F_l^+,l\in\{1,2,\dots\}.
\end{eqnarray}

Safeguarded constraints (\ref{SG1})-(\ref{SG2}) ensure, even after fixing the first $(t-h)\geq 0$ steps, the cumulative reward over each block never falls below  $(1-\varepsilon)$ of its optimum $F^{+}$. Note that these block-wise constraints are only activated when iteration $t$ resides in the block, and remains in subsequent MILP until the rolling window completely slides out of the block. In other words, constraints (\ref{SG1}) and (\ref{SG3}) can be  explicitly written as 
\begin{eqnarray}\label{SG3}
	\textstyle\sum_{h=0}^{t} f_{\xi_t}(x_h)+\sum_{h=t}^{\beta-1} f_{\xi_t}(x_h)
	\geq
	(1-\varepsilon)F_0^+,
\end{eqnarray}
\begin{eqnarray}\label{SG4}
	\textstyle\sum_{h=\beta+(l-1)\mu}^{t} f_{\xi_t}(x_h)+\sum_{h=t}^{\beta+l\mu-1} f_{\xi_t}(x_h)
	\geq
	(1-\varepsilon)F_l^+.
\end{eqnarray}

For both constraints (\ref{SG3}) and (\ref{SG4}), the first term on the left hand side denotes a fixed cumulative single step reward before time step $t$ within the block, and the second term represents the cumulative reward dependent on the current iteration decision. Next, Algorithm \ref{ALG} is presented  to illustrate the overall solution procedure of the safeguarded rolling process. Note that compared to traditional rolling framework, the safeguarded version introduces the segment concept, facilitating imposing reward bounds on iterations. If we consider the look-ahead window in traditional rolling optimization as an inner myopic ``preventer" compared to one-shot MILP, the safeguard introduces segment that is also for looking-ahead purposes but serving as an outer layer myopic ``perventer" across iterations. In this sense, safeguarded approach generalizes the traditional approach. If the segment length $\mu=1$, the safeguard constraints are trivially satisfied and reduces to traditional rolling optimization.

\begin{figure}[t]
	\centering
	\includegraphics[width=3.4in]{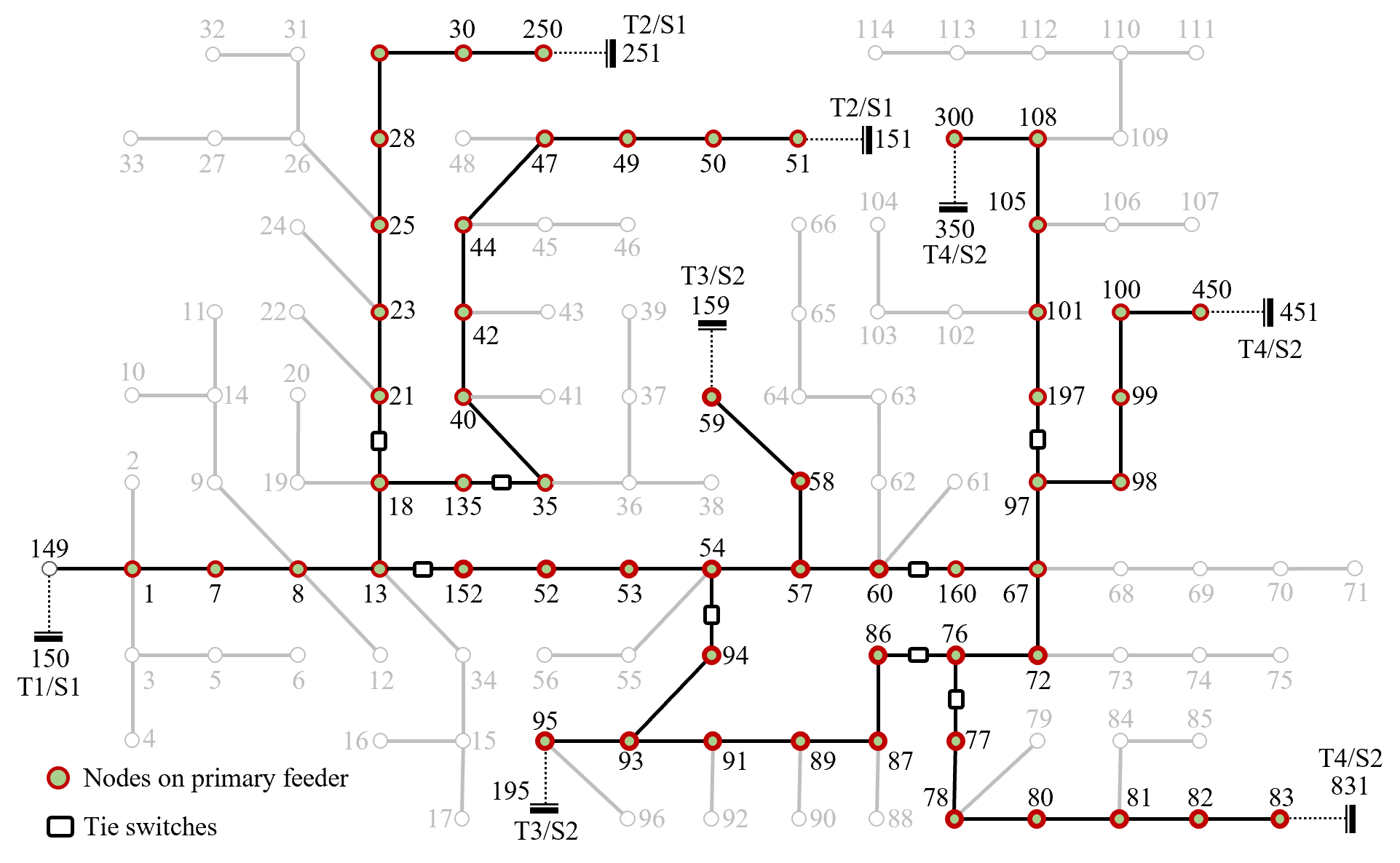}
	% where an .eps filename suffix will be assumed under latex, 
	% and a .pdf suffix will be assumed for pdflatex; or what has been declared
	% via \DeclareGraphicsExtensions.
	%\vspace{-5pt}
	\caption{An illustration of the modified IEEE 123-node test feeder.}
	\label{Testfeeder}\vspace{-5pt}
\end{figure} 

\vspace{-5pt}

\section{Case Study}

In this section, a detailed case study is conducted based on a modified IEEE 123-node distribution feeder. We first examine the performance of the proposed situationally aware multi-tier load restoration framework under both single and multiple fault scenarios to illustrate its dynamic multi-tier coordination. To further demonstrate its effectiveness, comparative analyses are performed against two benchmark strategies: a baseline one-shot MILP and a traditional rolling horizon approach.

\vspace{-5pt}

\subsection{Test System Setup}

The PDS is constructed based on a modified IEEE 123-node test feeder. The network totally consists of eight $4.16$ kV primary feeders supplied by two substations, each equipped with two transformers%, as shown in Fig. \ref*{Testfeeder}
. At substation S1, transformer T1 energizes the feeder at source node 150, while T2 serves those at nodes 251 and 151. Similarly, at substation S2,  transformer T3 connects the feeders at nodes 159 and 195, and T4 supplies nodes 831, 350 and 451. To simplify the network representation, all laterals are collapsed into their respective parent nodes on the main feeder. To support multi-tier load transfers, sectionalizing switches are installed between adjacent primary nodes, and a total of eight tie switches are deployed  at the end points of the feeders. Accordingly, the system is divided into 56 load blocks, with peak demands follow the standard IEEE 123-node feeder data. Upon re-energization, each load block exhibits a  demand surge, modeled with a CLPU factor of 2. Furthermore, 20\% of the load blocks are installed with behind-the-meter DERs, which can offset up to 50\% of their net demand after a random  delay, beginning no earlier than 6 steps. Transformer capacities are scaled relative to their loads: T1 has a capacity of 1.2 times its nominal load, T2 and T3 are sized at 1.5 times, and T4 is at 2 times, corresponding to utilization rates of approximately 80.0\%, 69.0\%, 73.6\%, and 55.5\%, respectively. The maximum power flow limit for each feeder is set to 5.0 times its nominal load. The look-ahead window length is set to 3 steps, with a maximum restoration horizon limited to 20 steps.

\vspace{-10pt}
\subsection{Evaluation of Multi-Tier Load Restoration}

\subsubsection{Single Fault Scenario}

\begin{figure}[t]
	\centering
	\includegraphics[width=3.2in]{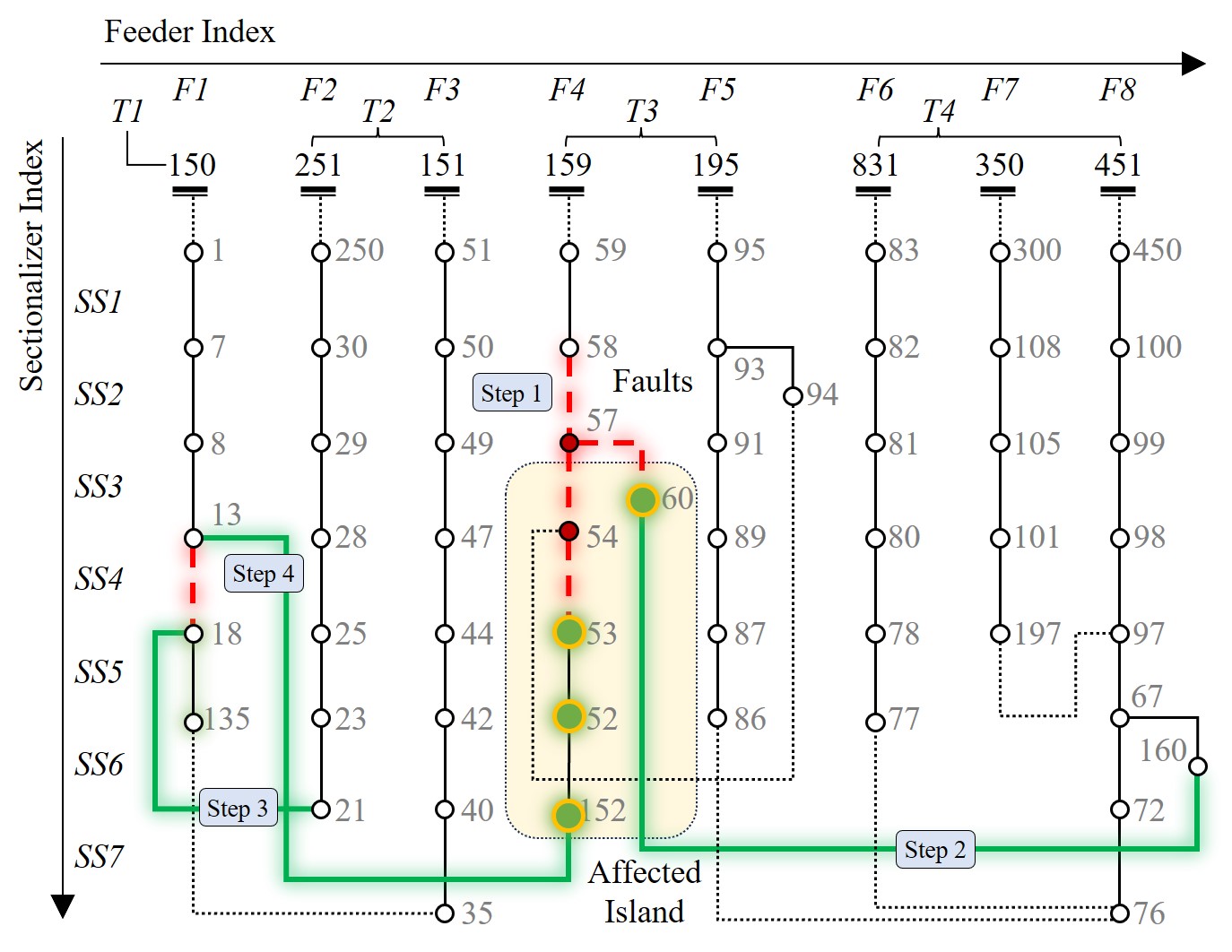}
	% where an .eps filename suffix will be assumed under latex, 
	% and a .pdf suffix will be assumed for pdflatex; or what has been declared
	% via \DeclareGraphicsExtensions.
	%\vspace{-5pt}
	\caption{Results of multi-tier restoration for single fault scenario.}
	\label{Singlefault}\vspace{-10pt}
\end{figure}

In the single fault scenario shown in Fig. \ref{Singlefault}, the multi-tier restoration requires five coordinated steps to complete a successful load transfer. The total restored load is 940.0 kW, with the utilization rates of transformers T1 through T4 being 93.3\%, 79.6\%, 27.3\%, and 92.6\%, respectively. At step 1, the three sectionalizing switches immediately upstream and downstream of the faulted area are opened for isolation. At step 2, the islanded load block containing node 60 is re-energized by closing the tie switch between feeders F4 and F8. Next, at step 3, the load blocks at nodes 18 and 135, originally supplied are transferred onto adjacent feeder F2 through sequential switching actions. Even though no load is restored in this step, it effectively frees up transformer T1’s capacity by 80.0 kW. This enables Step 4, where the affected island containing nodes 152, 52, and 53 can be successfully re-energized through back-feeding from feeder F1. Overall, the strategy utilizes up to tier-2 depth, achieving a total transferred load of 1040.0 kW. By contrast, if only adjacent feeders are considered without considering distant feeders, node 53 cannot be restored due to the insufficient capacity on transformer T1. Accordingly, the total restored load reduces to 780.0 kW, and the corresponding transformer utilization rates are reduced to 80.0\% and 69.0\% for T1 and T2, respectively.

\subsubsection{Multiple Fault Scenario}

\begin{figure*}[t]
	\centering
	\includegraphics[width=6.8in]{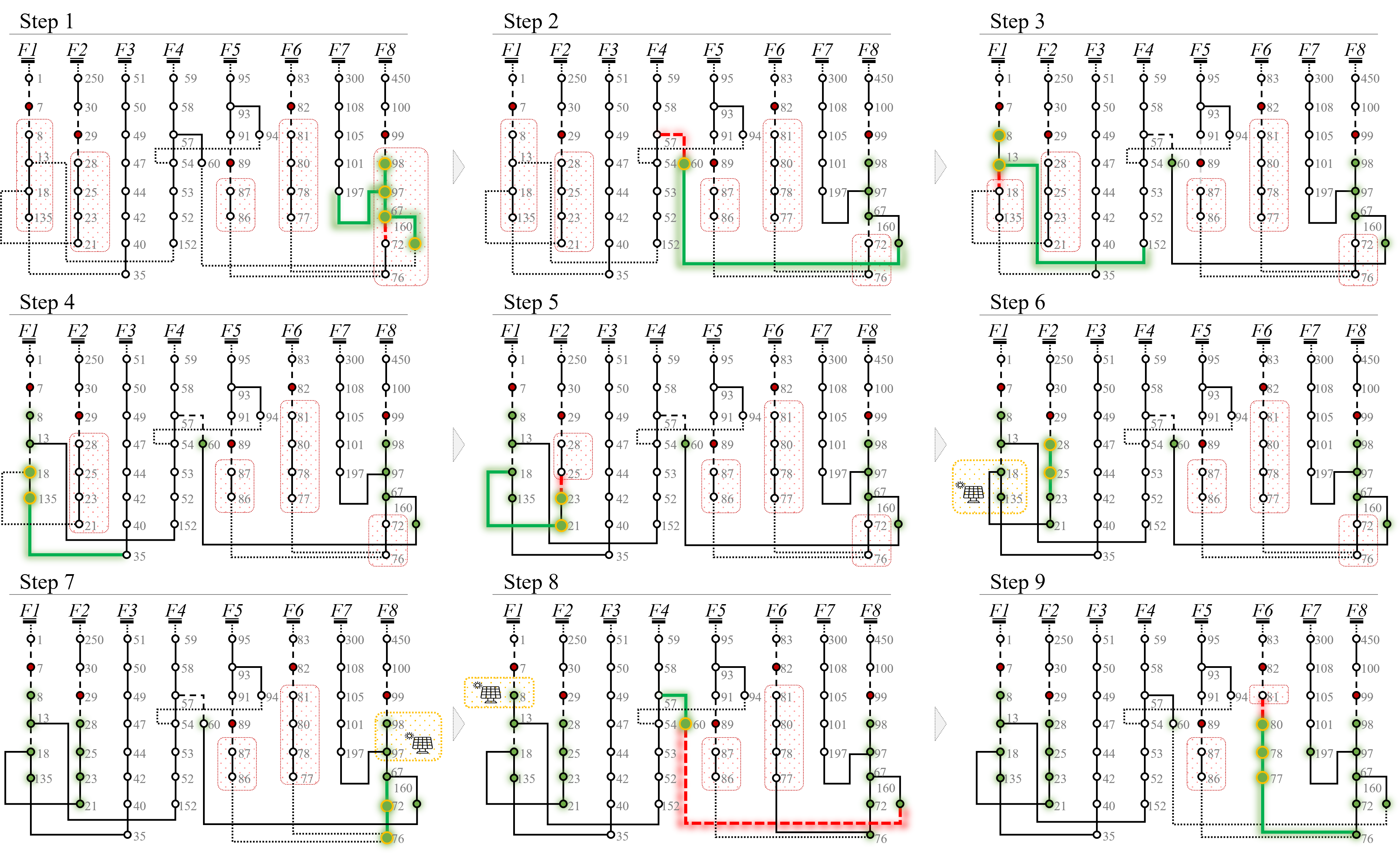}
	% where an .eps filename suffix will be assumed under latex, 
	% and a .pdf suffix will be assumed for pdflatex; or what has been declared
	% via \DeclareGraphicsExtensions.
	%\vspace{-5pt}
	\caption{An illustration of multi-tier restoration process considering major faults.}
	\label{Multifaults}\vspace{-5pt}
\end{figure*}

In this scenario, there are five faults occur across the distribution network, and the restoration process is illustrated in Fig.~\ref{Multifaults}, with several key steps offering insight into the multi-tier load transfer. In Step 2, the tie switch between feeders F4 and F8 is closed, thereby freeing 390.0 kW of capacity on transformer T3. This step enables the additional load at nodes 8 and 13 to be accommodated in Step 3. In Step 5, only a partial set of unserved load blocks (nodes 21 and 23) on feeder F2 is transferred to F3, as transferring all islanded blocks would raise the feeder load to 1515.0 kW, overloading transformer T2. From Step 6 onward, the benefits of situational awareness become evident. AMI data reveals that behind-the-meter DERs at nodes 18 and 135 have reconnected, reducing the net load on feeder F3 by 40.0 kW. This reduction enables the safe restoration of the remaining unserved blocks 25 and 28 at step 6. In subsequent steps, additional DER reconnections further alleviate the loading conditions on transformers T3 and T4, bringing their utilization rates below 85\%. Leveraging this real-time information, the operator reopens the tie between F4 and F8, returning block 60 to its original feeder F4, and then perform switch operations on F6 in Step 9 to energize blocks 77, 78 and 80. Finally, as all DERs have fully  reconnected, no improvement can be gained through further switching actions. The final outcome is 2370.0 kW of restored load, facilitated by a total load transfer of 3150.0 kW by using feeder connections extending up to tier-3 depth across the network.

\begin{figure}[t]
	\centering
	\includegraphics[width=3.4in]{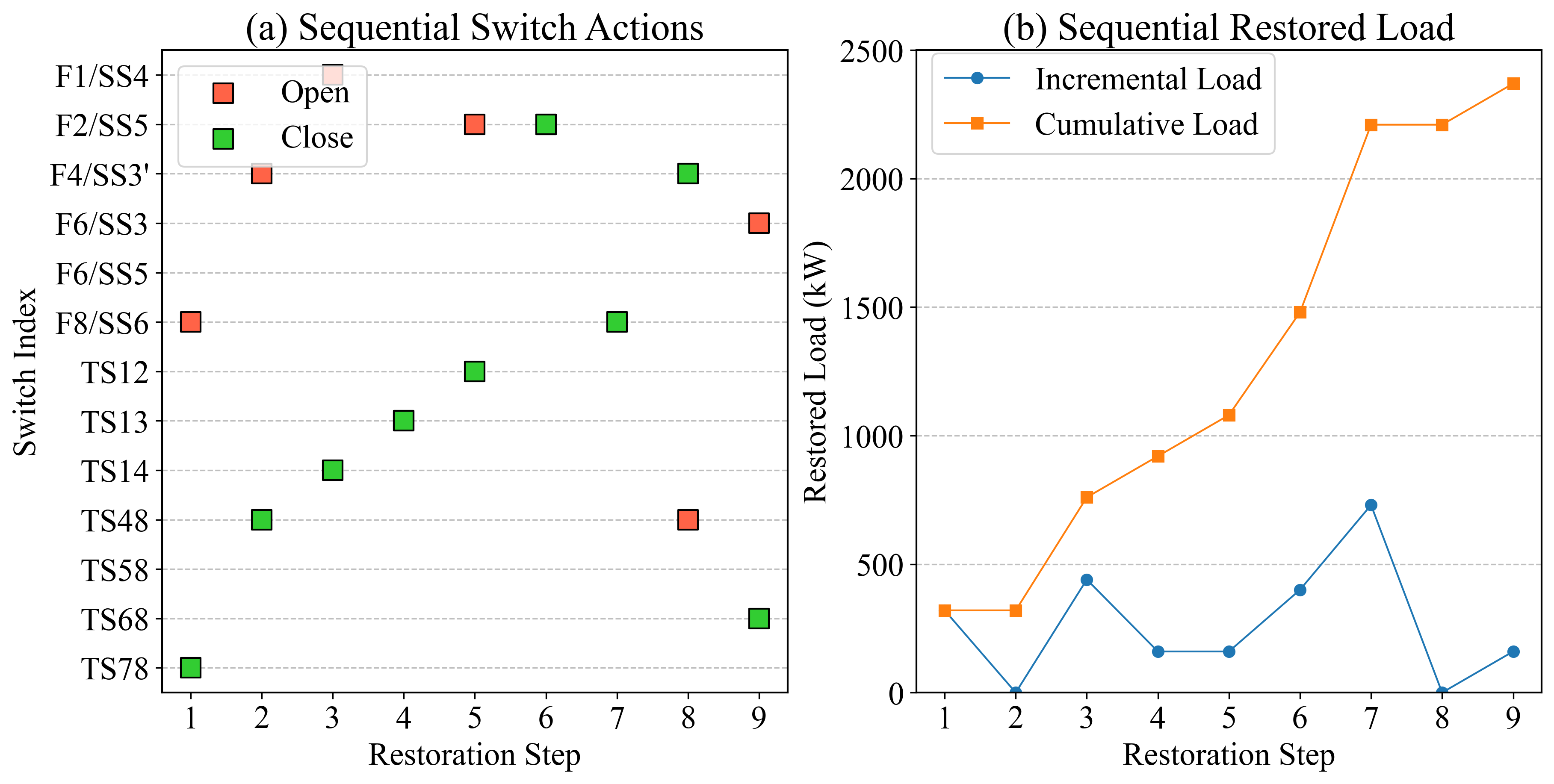}
	% where an .eps filename suffix will be assumed under latex, 
	% and a .pdf suffix will be assumed for pdflatex; or what has been declared
	% via \DeclareGraphicsExtensions.
	%\vspace{-5pt}
	\caption{Results of switch actions and restored load.}
	\label{Multiswitch}\vspace{-5pt}
\end{figure}

\begin{figure}[t]
	\centering
	\includegraphics[width=3.4in]{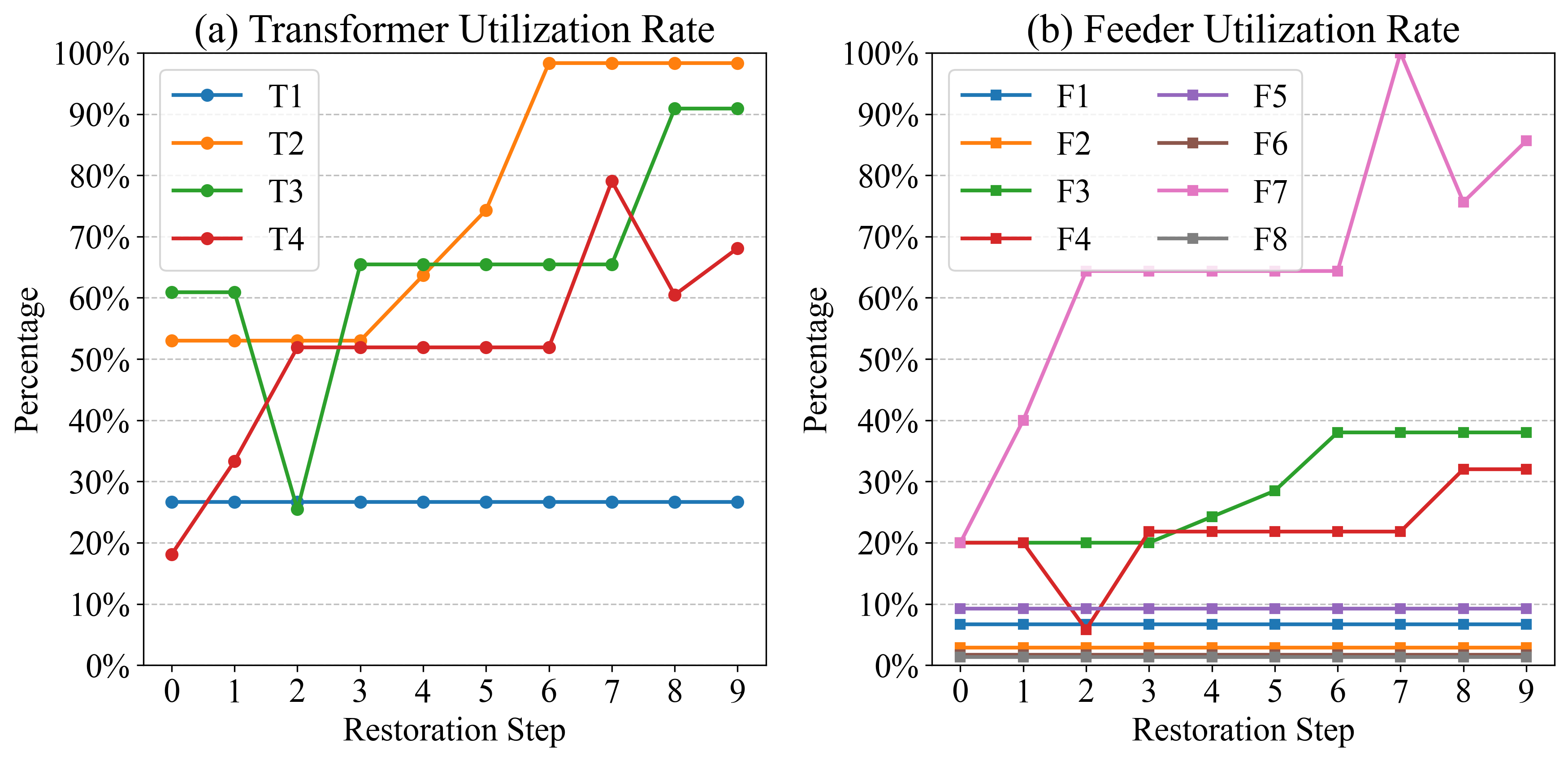}
	% where an .eps filename suffix will be assumed under latex, 
	% and a .pdf suffix will be assumed for pdflatex; or what has been declared
	% via \DeclareGraphicsExtensions.
	%\vspace{-5pt}
	\caption{Results of transformer and feeder utilization rate.}
	\label{Loading}\vspace{-5pt}
\end{figure}

Fig. \ref{Multiswitch}(a) shows switching actions executed at each step, with the related increases in restored load depicted in Fig. \ref{Multiswitch}(b). Fig. \ref{Loading} shows   transformer and feeder loading conditions. We can see that as the restoration progresses, the amount of restored load increases gradually, while transformer and feeder loadings initially rise fast and then tend to level off in later steps. The reason is load blocks originally supplied by transformer T1 via feeder F1 is sequentially redistributed to others. Yet, as DERs begin to reconnect, the incremental impact of load transfer on system loading alleviates. In addition, even though the network is congested due to load reallocation among healthy feeders, all transformer and feeder utilization remain  within limits.

Compared with the traditional rolling horizon, the proposed safeguarded approach effectively enhances the restoration performance. In step 1, the traditional approach will  immediately use the capacity of transformer T4 to energize all the  islanded load blocks on feeder F8. As a result, in subsequent steps, it can no longer back-feed node 60 through F7 nor shift blocks 8 and 13 onto F4. It means the restoration trajectory is locked in, restoring only 1820.0 kW within nine steps. By contrast, our safeguarded approach postpones the pickup of load blocks 72 and 76 until step 7. By deferring these actions, the proposed safeguarded approach reserves transformer T4 capacity during the early stages, thereby  maintaining the feasibility of deeper tier load transfers among feeders F1, F4, and F8.
This strategy not only enable more effective multi-tier coordination, but also leverage situational awareness by exploiting the potential DER reconnection that will be observed in future restoration steps.

\vspace{-10pt}

\subsection{Comparative Analysis and Performance Assessment}

To evaluate average performance, we compare three restoration strategies over 500 random simulation trials, each containing up to five randomly generated faults: (i) a baseline one-shot MILP approach without situational awareness, (ii) a traditional rolling horizon approach, and (iii) the proposed safeguarded rolling horizon approach. Table \ref{TRPM} summarizes the simulation results. The one-shot MILP approach yields the lowest average restored load of 1116.7 kW, because it cannot incorporate real-time AMI feedback and must rely on conservative peak load forecasts. Introducing a rolling horizon significantly improves performance, increasing the mean restored load to 1541.9 kW. However, it requires an average of 10.2 switching steps, and fails to converge within the first 20-step limits in 5.7\% of the trials. The degradations are primarily attributed to the myopic decisions made during early steps, which prematurely exhaust transformer capacity, leaving no feasible solutions for restoring more loads in later steps. In contrast, the proposed safeguarded rolling approach achieves the highest average restored load of 1869.2 kW, which represents a 21\% improvement over the traditional rolling approach, while reducing the average required steps to 7.3. These results underscore two advantages. One is that, by imposing segment-level restoration reward bounds, the safeguarded approach  preserves transformer capacity, enabling flexible load transfer across distant feeders and deferring some restoration actions to await the DER reconnections. Another is that by mitigating over-commitment of myopic decisions, the safeguard constraints reduces the need for corrective switching actions, effectively lowering the average step count by 28\%.

\begin{table}[t]
	\centering
	\begin{threeparttable}
		\caption{Evaluation of Average Restoration Performance}
		\label{TRPM}
		\begin{tabular*}{\linewidth}{@{\extracolsep\fill}l c c c@{}}
			\toprule
			Metric & One‐Shot & Traditional & Safeguarded \\
			\midrule
			\multirow{3}{*}{Restored load (kW)}
			& [947.2]\tnote{1}   & [1241.0] & [1643.1] \\
			& 1116.7             & 1541.9   & 1869.2   \\
			& (1297.5)\tnote{1}  & (1765.7) & (2121.5) \\
			\addlinespace
			\multirow{3}{*}{Number of steps}
			&    & [7.5]  & [5.2]  \\
			& N/A     & 10.2   & 7.3  \\
			&    & (12.9) & (8.9)  \\
			
			\addlinespace
			Failure rate ($>20$ steps)\tnote{2}
			& N/A      & 5.7\%     & 0.2\%      \\
			\bottomrule
		\end{tabular*}
		\begin{tablenotes}[flushleft]
			\item[1] [ ] = minus one standard deviation; ( ) = plus one standard deviation.
			\item[2] Percentage of simulation trials that did not converge within 20 steps.
		\end{tablenotes}
	\end{threeparttable}
\end{table}

\begin{figure}[t]
	\centering
	\includegraphics[width=3.4in]{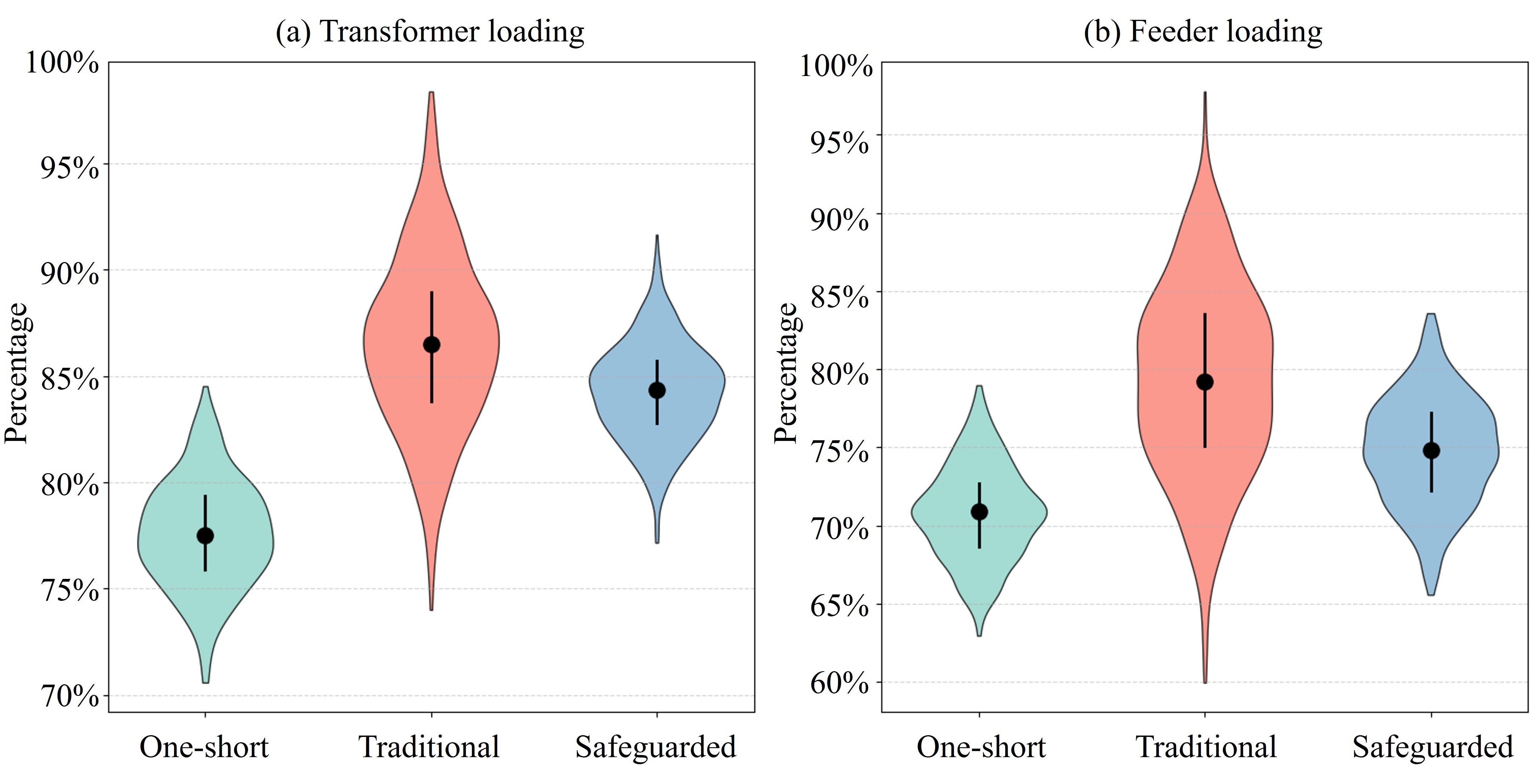}
	% where an .eps filename suffix will be assumed under latex, 
	% and a .pdf suffix will be assumed for pdflatex; or what has been declared
	% via \DeclareGraphicsExtensions.
	%\vspace{-5pt}
	\caption{Statistic analysis of network loading conditions.}
	\label{CPRS}\vspace{-5pt}
\end{figure}

Fig. \ref{CPRS} depicts the maximum loading statistics over 500 simulation trials. The one-shot MILP approach exhibits the lowest mean transformer utilization rate, primarily because it is solved only once and lacks the ability to adapt to the evolving system conditions, such as actual net demand and DER reconnection. By employing rolling optimization framework, the utilization of transformers is improved. However, the large standard deviation observed under the traditional rolling approach indicates significant imbalance in load transfer. In other words, in many trials, one or two transformers are operated near their thermal limits, while others remain lightly loaded. This imbalance is also obvious in feeder loading conditions. As plotted in Fig. \ref{CPRS}, the maximum per-trial feeder rate under the traditional rolling approach exhibits not only a higher average but also a heavier tail. By contrast, the proposed safeguarded  approach improves the transformer utilization,  while simultaneously reducing its standard deviation. The distribution of maximum feeder rate is compressed toward lower values. These results validate that the safeguard constraints promote more effective and balanced multi-tier load transfers.

\vspace{-10pt}

\section{Conclusion}

This paper proposes a situationally aware rolling horizon multi-tier load restoration framework that coordinates substation transformer and feeder capacities while accounting for behind-the-meter DER dynamics and CLPU effects. A multi-tier load transfer model is developed, capturing the progressive switching and load reallocation process using binary actional variables, enabling restoration beyond adjacent feeders. To address the delayed DER reconnection issue, the problem is formulated as an MILP and embedded within a rolling horizon optimization framework. A safeguarded rolling strategy is proposed to mitigate myopic decisions by enforcing segment-level reward bounds. Simulations on a modified IEEE 123-node test feeder demonstrate that the proposed framework significantly enhances load restoration performance and system utilization compared to one-shot MILP and traditional rolling approaches. Future research can investigate the scalability of the multi-tier restoration, as step by step execution increases the number of binary variables rapidly. AI-augmented optimization provides a promising direction to manage this complexity.

\bibliographystyle{IEEEtran}\bibliography{ref}

\end{document}